\date{}
\documentclass[11pt]{article}
\usepackage{amsmath,amsthm,amsfonts,amssymb,amsopn,amscd}     
\usepackage{color}

\def\tilde{\widetilde}

\def\cL{\cal L}

\def\A{{\cal A}}
\def\B{{\cal B}}

\def\H{{\cal H}}
\def\K{{\cal K}}
\def\S{{\cal S}}

\def\cU{{\cal U}}

\def\PSL{{{\rm PSL}(2,\mathbb R)}}

\def\S2{S^{1(2)}}
\def\Poi{{\cal P}_+^\uparrow}
\def\tPoi{\tilde{\cal P}_+^\uparrow}
\newtheorem{theorem}{Theorem}[section]
\newtheorem{lemma}[theorem]{Lemma}

\newtheorem{corollary}[theorem]{Corollary}

\newtheorem{proposition}[theorem]{Proposition}

\theoremstyle{definition} 

\theoremstyle{remark} 

\def\PSL{PSU(1,1)}

\def\RR{{\mathbb R}}
\def\CC{{\mathbb C}}

\def\SL2{{{\rm SL}(2,\RR)}}
\def\SLC{{{\rm SL}(2,\CC)}}
\def\PSL2{{{\rm PSL}(2,\RR)}}

\def\U1{{{\rm V}(1)}}
\def\SU2{{{\rm SV}(2)}}

\def\SU{{{\rm SU}}}

\def\A{{\mathcal A}}
\def\B{{\mathcal B}}

\def\H{{\mathcal H}}

\def\K{{\mathcal K}}

\def\cO{{\mathcal O}}

\def\cP{{\mathcal P}}

\def\W{{\mathcal W}}

\def\polhk#1{\setbox0=\hbox{#1}{\ooalign{\hidewidth
    \lower1.5ex\hbox{`}\hidewidth\crcr\unhbox0}}}

\title{\Huge An algebraic condition for the Bisognano-Wichmann Property}
\author{{\sc Vincenzo Morinelli\footnote{partially supported by ERC and GNAMPA-INdAM} }
\\
Dipartimento di Matematica,
Universit\`a di Roma Tor Vergata,\\
Via della Ricerca Scientifica, 1, I-00133 Roma, Italy\\
E-mail: {\tt morinell@mat.uniroma2.it}
}
\begin{document}

\maketitle

\begin{abstract}
The Bisognano-Wichmann property for  local, Poincar\'e covariant nets of standard subspaces is discussed.  We present a sufficient algebraic condition on the covariant representation ensuring  Bisognano-Wichmann and Duality properties  without further assumptions on the net. Our {\it``modularity"} condition holds for direct integrals of scalar massive and masselss representations. We conclude that in these cases the Bisognano-Wichmann property is much weaker than the Split property. Furthermore, we present a class of massive modular covariant nets not satisfying the Bisognano-Wichmann property.
\end{abstract}


\section{Introduction}

\indent The Tomita-Takesaki theory for von Neumann algebras  has a crucial role in the study of geometric properties of Quantum Field Theories. Bisognano and Wichmann  showed that, in Wightman theories, modular groups associated to wedge regions have a geometrical interpretation: they implement pure Lorentz transformations \cite{BW1,bw2}.

Let $\RR^{1+3}\supset\cO\mapsto \A(\cO)\subset \B(\H)$ be a local Poincar\'e covariant net of von Neumann algebras on a fixed Hilbert space $\H$, from certain open, connected bounded regions of the Minkowski spacetime $\mathbb R^{1+3}$. Let $U$ be a  positive energy Poincar\'e representation acting covariantly on $\A$ and let $\Omega$ be the unique (up to a phase) normalized $U$-invariant vector, namely the vacuum vector. $\Omega$ is supposed to be cyclic for all the local algebras.  The \textbf{Bisognano-Wichmann}  property (B-W) is stated as follows:
\begin{equation}
\label{eq:biwi}
U(\Lambda_W(2\pi t))=\Delta_{\A(W),\Omega}^{-it},\end{equation}
for any wedge shaped region $W$, where $t\mapsto\Lambda_{W}(t)$ is the one parameter group of boosts associated to the wedge $W$ and $\Delta^{it}_{\A(W),\Omega}$ is the modular group associated to $\A(W)=\left(\cup_{\cO\subset W}\A(\cO)\right)''$ w.r.t. $\Omega$.
The Bisognano-Wichmann property strictly links the covariant representation of the Poincar\'e group to the net: in a precise sense it is enclosed in the net once we choose the vacuum state and the algebra of the observables. 

It is a basic fact in many aspects of QFT: it holds in massive theories \cite{M}, it implies the Spin-Statistics canonical relations \cite{GL} and the essential Duality property.  References on Duality and Bisognano-Wichmann properties can be found in Refs.~\cite{bor,Bo2,sch,BDFS,M,B}. In general, counter-examples have a pathological nature. For instance, theories  having an infinite number of particles in the same mass multiplet do not need to satisfy the B-W property (cf. Sect. \ref{sect:CO}).

The Bisognano-Wichmann property cares only about the modular structure of the algebraic model which is contained in the standard subspace structure of the net. 
Standard subspace techniques give a fruitful approach to QFT. For instance, they are useful to model first quantization nets, giving a canonical structure to free fields (cf. Brunetti-Guido-Longo construction in Ref.~\cite{BGL}), and they have a keyrole in finding localization properties of Infinite Spin particles, cf. Ref. \cite{LMR}.

One should expect an algebraic description of the B-W property. We ask  when a (unitary, postive energy) Poincar\'e representation $U$ is {\it modular}, i.e. any net of standard subspaces it acts covariantly on satisfies the B-W property, in particular $U$ is implemented by modular operators. This is expected, at least, for irreducible Poincar\'e, positive energy representations.
 We  present  an {\it algebraic condition} ensuring the modularity of a representation. It holds  at least for scalar Poincar\'e representations in any spacetime $\mathbb R^{1+s}$, with $s\geq3$. 

This analysis is useful to compute counterexamples. We describe a massive, modular covariant net, not respecting the B-W property. The massless case is contained in Ref.~\cite{LMR}. 
Furthermore, an  outlook  on the relation between modular covariance and Split property on generalized free fields is given. Details about this analysis can be found in Ref.~\cite{Mo}.

\section{One particle net}
A linear, real, closed subspace $H$ of a complex Hilbert space $\H$ is called \emph{cyclic} if 
$H+iH$ is dense in $\H$, \emph{separating} if $H\cap iH=\{0\}$ and 
\emph{standard} if it is cyclic and separating.
Let $H$ be a standard subspace, the   Tomita operator $S_H$ is  the closure, of the densely defined anti-linear involution
$$ H+iH\ni \xi + i\eta \mapsto \xi - i\eta\in H+iH, \qquad\xi,\eta\in H.$$ Its polar decomposition $S_H = J_H\Delta_H^{1/2}$ defines the positive selfadjoint \emph{modular operator} $\Delta_H$ and the anti-unitary
\emph{modular conjugation} $J_H$. $\Delta_H$ is invertible and \begin{equation}\label{tom}J_H\Delta_H J_H=\Delta_H^{-1}.\end{equation} 
Standard subspaces are in 1-1 correspondence with densely defined antilinear involutions  which are in 1-1 correspondence with pairs of an antiunitary, involutions and selfadjoint operators satisfying \eqref{tom}.
Here is a basic Lemma on standard subspaces.
\begin{lemma}\label{lem:sym}{\rm \cite{L,LN}}
Let $H,K\subset\H$ be standard subspaces  and $U\in\cU(\H)$ be a unitary operator on $\H$ s.t. $UH=K$. Then $U\Delta_H U^*=\Delta_K$ and $UJ_HU^*=J_K$.
\end{lemma}

Consider the (proper, orthocronous) Poincar\'e group $\Poi$ of the 3+1 dimensional Minkowski spacetime $\RR^{3+1}$. $\Poi$ is the semidirect product of 4-translations and the connected component of the identity of the Lorentz group denoted by $\cL_+^\uparrow$. 
A wedge shaped region $W$ in $\RR^{3+1}$ is an open region of the form  $gW_1$ where $g\in\Poi$ and $W_1=\{p\in\RR^{3+1}: |p_0|<p_1\}$. $W'$ denotes the causal complement of $W$ w.r.t. the causal structure of $\RR^{3+1}$.
The set of wedges is  denoted by $\W$. $\W_0$ denotes the subset of wedges of the form $gW_1$ where $g\in \cL_+^\uparrow$. For every wedge $W\in\W$ there exists a unique 1-parameter group of Poincar\'e boosts $t\mapsto\Lambda_W$ preserving $W$, i.e. $\Lambda_W(t) W=W$ for every $t\in\RR$.

Let $U$ be a unitary, positive energy representation (p.e.r.) of the Poincar\'e group $\Poi$ on  an Hilbert space $\H$. 
We shall call a $U$-covariant net of standard subspaces on wedges  a map $$H:\W\ni W\longmapsto H(W)\subset\H,$$
associating to every wedge in $\RR^{1+3}$ a closed real linear subspace of $\H$, s.t.:
 \begin{itemize}
\item[1.] \textbf{Isotony:}  if $W_1,W_2\in\W$, $W_1\subset W_2$ then $H(W_1)\subset H(W_2)$
\item[2.] \textbf{Poincar\'e Covariance:} $\forall W\in\W$, $g\in\Poi$: $U(g)H(W)=H(gW).$
\item[3.] \textbf{Reeh-Schlieder Property:}  $\forall W\in\W$, $H(W)$ is a cyclic subspace of $\H$.
\item[4.] \textbf{Locality:}   $\forall W_1,W_2\in\W$ s.t. $W_1\subset W_2'$: $  H(W_1)\subset H(W_2)'.$\end{itemize} 

We shall indicate with $(U,H)$ a Poincar\'e covariant net of standard subspaces satisfying 1-4\footnote{It is possible to generalize the assumptions to fermionic representations assuming twisted-locality on $H$ \cite{LMR}. The  forthcoming analysis continues to hold.}.
The \textbf{Bisognano-Wichmann Property} in this setting is stated as follows:
$$U(\Lambda_W(2\pi t))=\Delta^{-it}_{H(W)},\qquad\forall t\in\RR.$$ {\bf Duality property} states that $H(W)=H(W')'$.

The proper Poincar\'e group $\cP_+$ is generated by $\Poi$ and the space and time reflection $\Theta$. An irreducible unitary representation of the Poincar\'e group always extends to an  (anti-)unitary representation of $\cP_+$ (i.e. unitary on $\Poi$, anti-unitary on $\theta\Poi$)  except for finite helicity representations which have to be coupled. The following theorem is a consequence of the Brunetti-Guido-Longo construction. 
\begin{theorem}\label{thm:bgl}\cite{BGL}
There is a 1-1 correspondence between:
\begin{itemize}
\item[a.] (Anti-)unitary positive energy representation of $\cP_+$.
\item[b.] Local nets of standard subspaces  satisfying 1-4 and B-W Property.
\end{itemize}
\end{theorem}
In view of Theorem \ref{thm:bgl}, our question becomes: which families of representations preserve the 1-1 correspondence when the B-W property is not assumed?

\section{A modularity condition for the Bisognano-Wichmann Property}

For every $W\in\W_0$, we shall call $G_W^0$ the set $\{g\in{\cal {L}_+^\uparrow}:  g W=W\}$. $G_W$ denotes the Poincar\'e subgroup generated by  the translation group  and $G_W^0$. By transitivity of $\Poi$ on wedges, $G_W^0$ and $G_W$ can be defined for any $W\in\W$.
The strongly continuous map $$Z_{H(W)}:\RR\ni t\mapsto \Delta_{H(W)}^{it}U(\Lambda_W(2\pi t))\in\cU(\H)$$ is the proper quotient between the modular group of a wedge subspace and the  unitary implementation of the one parameter boost transformations associated to $W$. Indeed, it has to be identity map if the B-W property holds.

The following proposition is a consequence of $\Poi$-covariance and Lemma \ref{lem:sym}. 
\begin{proposition}\label{prop:sym} 
Let $(U,H)$ be a Poincar\'e covariant net of standard subspaces. Then, for every $W\in\W$, $t\mapsto Z_{H(W)}(t)$
defines a one-parameter group and
 $$Z_{H(W)}(t)\in U(G_W)',\qquad \forall t\in\RR.$$
 \end{proposition}

Here is stated our algebraic condition, sufficient to modularity.\\
\noindent {\bf Modularity condition}: let $r_W\in\Poi$, s.t. $r_WW=W'$,  then
\begin{equation}\tag{M}\label{eq:cond}{U(r_W)\in U(G_W)''}.\end{equation}
\begin{theorem}\label{teo:then} Let $(U,H)$ be a Poincar\'e covariant net of standard subspaces.
Let $W\in\W$, and $r_W\in\Poi$ be s.t. $r_WW=W'$. If $Z_{H(W)}$ commutes with $U(r_W)$, then the  Bisognano-Wichmann and Duality properties hold.
\end{theorem}
The proof goes as follows. By hypotheses and covariance we have $Z_{H(W)}=Z_{H(W')}$. As $Z_{H(W)}$ is a one parameter group of automorphisms of $H(W)$, by the KMS condition,   Duality and B-W properties follow.

It is a corollary that representations  satisfying $\eqref{eq:cond}$ are modular.
\begin{corollary}\label{starcor}
With the assumptions of Theorem \ref{teo:then}, if condition \eqref{eq:cond} holds on $U$, then Bisognano-Wichmann  and Duality properties hold on $H$.
\end{corollary}
The above property easily extends to direct integrals and multiples of representations as the following proposition shows.
\begin{proposition}\label{prop:P} Let $U$  and $\{U_x\}_{x\in X}$ be p.e.r of $\tPoi$  satisfying $(M)$.

Let $\K$ be an Hilbert space, $(M)$ holds for $U\otimes 1_\K\in\B(\H\otimes \K)$.

Let $(X,\mu)$ be a standard measure space.
Assume that  $U_x|_{G_W}$ and $U_y|_{G_W}$ are disjoint for $\mu$-a.e. $ x\neq y$ in $X$. Then $U=\int_X U_x d\mu(x)$ satisfies $(M)$. \end{proposition}

\section{The scalar case}
As we have announced, the above analysis holds for scalar representations. 

\begin{proposition}\label{prop:scalar} Let $U$ be a unitary irreducible scalar representation of the Poicar\'e group. Then $U$ satisfies the modularity condition \eqref{eq:cond}.
\end{proposition}
The proof relies on two facts. Translations of a scalar representation generate a maximal abelian sub-algebra (MASA) of $\B(\H)$  and $r_W$ preserves $G_W^0$ orbits of the $m$-mass hyperboloid (a.e. orbits in the massless case).

\begin{theorem}\label{starint}Let $U=\int_{[0,+\infty)}{U_m}d\mu(m)$ where $\{U_m\}$ are (finite or infinite) multiples of the scalar representation of mass $m$.  Then $U$ satisfies $(M)$, hence Duality and Bisognano-Wichmann properties hold.\end{theorem} 
It is a consequence of Theorem \ref{prop:scalar} and Proposition \ref{prop:P}.

 This analysis holds in any Minkowski spacetime $\RR^{s+1}$ with $s\geq3$.

\section{Counterexamples and Outlook}\label{sect:CO}

Consider $U_{m,s}$ $m$-mass, and $s$-spin, unitary, irreducible representation of the  Poincar\'e group $\Poi$. Let $ W\mapsto H(W)$ be the canonical net associated to $U_{m,s}$.
Let $V$ be a unitary, non trivial, representation of ${\cL}_+^\uparrow$ on an Hilbert space $\K$, which is real, i.e. there exists  a standard subspace $K\subset\K$ preserved by $V$ and s.t.  $ \Delta_K=1$ and $J_K K=K$.
We can define the following new net of standard subspaces (cf. Ref.\cite{LMR} for standardness), $$K\otimes H:\W\ni W\longmapsto K\otimes H(W) \subset\K\otimes \mathcal{H}.$$  
Two Poincar\'e representations act covariantly on $K\otimes H$: 
$$U_I(a,\Lambda)\equiv1\otimes U_{m,s}(a,\Lambda)
\qquad \textrm{and}
\qquad 
U_V(a,\Lambda)\equiv V(\Lambda)\otimes U_{m,s}(a,\Lambda),$$
where
$\Lambda\in {\cL}_+^\uparrow,\, a\in\RR^{3+1}$. 
$U_I$ is implemented by $K\otimes H$ modular operators and Bisognano-Wichmann property holds w.r.t. $U_I$.   Note that when  $s=0$, $U_I$ satisfies the modularity condition (\ref{eq:cond}), hence the B-W property  also follows by Corollary \ref{starcor}. It is obvious by construction that $U_V$ does not satisfy B-W property on $K\otimes H$ except when $V$ is the trivial representation. Notice that $V$ could be chosen as a representation of $\SLC$ (i.e.\! the universal covering of $\cL_+^\uparrow$) and thus a wrong Spin-Statistics relation would follow for $U_V$.

$U_V $ decomposes into a direct sum of infinitely many inequivalent representations of mass $m$, in particular infinitely many spins have finite multiplicity \cite{Mo}. It is interesting to look for natural counterexamples to modular covariance, in the class of representations excluded by this discussion, if they exist.  We expect that a modularity condition could be established on a finite direct sum of factorial Poincar\'e representations.

We briefly give an outlook on the relation between B-W and Split property, a strong statistical independence property in QFT defined by Doplicher and Longo in Ref.~\cite{lodo}. 

An inclusion of von Neumann algebras $(N\subset M,\Omega)$ is said to be {\it Split} if there exists an intermediate type I factor $F$  $(N\subset F\subset M)$. 
Analogously we define an inclusion of standard subspaces $K\subset H$ {\it Split} iff their second quantization von Neumann algebras give a Split inclusion. 
Our analysis suggests that the  Bisognano-Wichmann property is {\it much  weaker} than the Split property. Indeed, consider a net $(U,H)$ be a Poincar\'e covariant net, assume that $U$ is a direct integral of scalar representations. The Split property on the $U$-canonical net requires that  $U$ shall be purely atomic on masses, concentrated on isolated points and for each mass there can only be a finite multiple of the scalar representation\cite{lodo,Mo}. The disintegration satisfies the modularity condition (M) and the modularity of $U$ follows by Corollary \ref{starcor}.

\section*{Acknowledgements} I hearthful thank Roberto Longo for suggesting me the problem and Mih\'aly Weiner for  comments on preliminary version of this paper.

\end{document}